\documentclass{article}
\makeatletter
\@addtoreset{equation}{section}

\makeatother

\begin{document}

\begin{flushright}
May 2006

KUNS-2023
\end{flushright}

\begin{center}

\vspace{5cm}

{\LARGE 
\begin{center}
CFT for Closed String Tachyon Condensation
\end{center}
}

\vspace{2cm}

Takao Suyama \footnote{e-mail address : suyama@gauge.scphys.kyoto-u.ac.jp}

\vspace{1cm}

{\it Department of Physics, Kyoto University,}

{\it Kitashirakawa, Kyoto 606-8502, Japan }

\vspace{4cm}

{\bf Abstract} 

\end{center}

We construct a class of 
CFT's which describe space-dependent closed string tachyon backgrounds, as the IR limit 
of GLSM's in which the FI-parameter is promoted to a superfield. 
Whole process of tachyon condensation is described by a single CFT. 
We apply this construction to several examples, in which target space is deformed drastically, and 
the dilaton background may vary, as a tachyon condenses.

\newpage

\vspace{1cm}

\section{Introduction}

\vspace{5mm}

Understanding of closed string tachyon condensation has been gradually proceeding. 
In recent researches, it has been recognized that there is a class of tachyons whose condensation can 
be understood rather easily. 
It is a class in which tachyons are localized in a non-compact space. 
A typical and well-studied example of such a tachyon is the one appearing in string theory on a 
non-compact and non-supersymmetric orbifold \cite{APS}. 

For the other tachyons, their condensations are studied recently in 
\cite{Suyama1}\cite{recent2}\cite{recent1}. 
One property which seems to be common in the latter tachyons is that they inevitably couples to the 
dilaton, and the endpoint of their condensations would have linear dilaton backgrounds 
\cite{YZ1}\cite{MIT}\cite{Suyama1}. 
The analysis of this phenomenon has been done by using a spacetime effective action which is valid 
when the size of string is negligible, but $\alpha'$-corrections are not always small, especially 
when the tachyon mass$^2$ is of order of $(\alpha')^{-1}$. 
There is an analysis by using string field theory \cite{YZ2}, and also by worldsheet RG flows 
\cite{Suyama2}. 

In this paper, we construct $N=2$ SCFT's which describe an $\alpha'$-exact background of string theory, 
in which a tachyon varies along a spatial direction. 
They are obtained as the IR limit of gauged linear sigma models (GLSM's) \cite{GLSM}. 
GLSM has been already used in \cite{Vafa} to discuss the non-supersymmetric orbifolds, in which GLSM 
provides a control of an RG flow which is assumed to describe the corresponding tachyon condensation. 
Our construction is, roughly speaking, to promote the RG scale to one of the target space coordinate, 
and to make the whole tachyon background on-shell. 
We show that there is such a CFT in which the dilaton gradient varies as the tachyon varies. 
This can be regarded as an explicit realization of the claim made in \cite{Suyama2}\cite{Suyama1}. 
Throughout this paper, we just focus on the tree level string theory, and possible problems on a 
strong coupling region will be, hopefully, discussed elsewhere. 

This paper is organized as follows. 
In section \ref{GLSM}, 
we briefly review \cite{Vafa}, and then explain how to obtain an on-shell background 
from the corresponding RG flow. 
Some applications of our construction to tachyon condensations are shown in section \ref{example}. 
The examples include vanishing of target space, and pinching off of a cylinder discussed in 
\cite{Silverstein}. 
Section \ref{W} deals with tachyon condensations in which the dilaton varies with tachyons. 
We comment on properties of our CFT in section \ref{charge}. 
Section \ref{discuss} is devoted to discussion. 
Our conventions of the superfields and the detailed construction of $N=2$ superconformal algebra, 
which is a review of 
\cite{SCA1}\cite{SCA2}\cite{HoriKapustin} 
in a general setup, are shown in appendices.

\vspace{1cm}

\section{On-shell tachyon condensation in GLSM} \label{GLSM}

\vspace{5mm}

\subsection{Tachyon condensation and RG flow}

\vspace{5mm}

First let us recall a description of a closed string tachyon condensation by a GLSM \cite{Vafa}. 
The GLSM is a $U(1)$ gauge theory in two dimensions with $(2,2)$ supersymmetry. 
Suppose that there are $n$ chiral superfields $\Phi_i$ whose charges for the $U(1)$ gauge symmetry 
are $q_i$. 
The Lagrangian of this GLSM is 
\begin{equation}
L = \frac1{2\pi}\int d^4\theta\ \Bigl( \sum_{i=1}^n\bar{\Phi}_ie^{2q_iV}\Phi_i
 -\frac1{2e^2}\bar{\Sigma}\Sigma \Bigr) -\frac1{2\pi}\Bigl[ \int d^2\tilde{\theta}\ t\Sigma 
 +\mbox{h.c.}\ \Bigr],  
    \label{Vafa}
\end{equation}
where $\Sigma=\bar{D}_+D_-V$ is a twisted chiral superfield constructed from $V$, and 
$t=t_R+it_I$ is a 
complex parameter. 
For conventions see Appendix \ref{convention}. 

The classical vacua is determined by the D-term potential. 
The minima of the potential is the solutions of 
\begin{equation}
\sum_{i=1}^nq_i|\phi_i|^2 = 2t_R, 
   \label{target}
\end{equation}
where $\phi_i$ are the lowest components of $\Phi_i$. 
By taking $e\to\infty$ limit, fluctuations perpendicular to this vacuum manifold (\ref{target}) 
become infinitely 
heavy, and the GLSM approaches a non-linear sigma model whose target space is the vacuum manifold 
(\ref{target}). 

To see the geometry of (\ref{target}) in more details, suppose that $q_1,\cdots,q_l$ are positive, 
and the rest of $q_i$ are negative. 
If $t_R$ is negative, then all $\phi_{l+1},\cdots,\phi_n$ cannot vanish simultaneously. 
These can be regarded as homogeneous coordinates of the weighted projective space 
$WCP_{q_{l+1},\cdots,q_n}$. 
The total space of (\ref{target}) is a non-compact bundle over this $WCP_{q_{l+1},\cdots,q_n}$, 
which is topologically equivalent to ${\bf C}^l\times({\bf C}^{n-l}\backslash\{\vec{0}\})/\sim$ where 
the identification is defined as follows, 
\begin{equation}
(\phi_1, \cdots, \phi_n) \sim (\lambda^{q_1}\phi_1,\cdots,\lambda^{q_n}\phi_n), \hspace{5mm} 
\lambda\in{\bf C}^*. 
\end{equation}
If $t_R$ is positive, then (\ref{target}) determines a similar non-compact bundle over 
$WCP_{q_{1},\cdots,q_l}$. 

Note that if all $q_i$ are positive (negative), then there is no solution to (\ref{target}) when 
$t_R$ is negative (positive), respectively. 
In those cases, there are no supersymmetric vacua for such $t_R$ regions. 

\vspace{5mm}

The GLSM is super-renormalizable, and thus the renormalization procedure is very simple. 
In fact, all divergences are canceled by the shift of $t_R$ as follows, 
\begin{equation}
t_R(\Lambda) = t_R+\frac12\sum_{i=1}^nq_i \log\Lambda, 
   \label{t-shift}
\end{equation}
where $\Lambda$ is a UV cut-off scale. 
If $\sum_iq_i>0$, then $t_R(\Lambda)$ decreases monotonically as the scale $\Lambda$ decreases, 
and vice versa. 
Recalling that the shape (even topology) of the vacuum manifold, in other words, the target space 
of the corresponding non-linear sigma model, depends on the sign of $t_R$, it is concluded that 
the target space varies drastically along the RG flow. 

To see what is described by this RG flow, it is convenient to dualize $\Phi_i$, according to 
\cite{HoriVafa}. 
The twisted superpotential of the dual theory is 
\begin{equation}
\tilde{W} = \Sigma\Bigl[ \frac12\sum_{i=1}^nq_iY_i-t \Bigr]+\mu\sum_{i=1}^ne^{-Y_i}, 
\end{equation}
where $Y_i$ are twisted chiral superfield dual to $\Phi_i$, and $\mu$ is a scale parameter related 
to $\Lambda$ and $t$. 
The superpotential is absent in the dual theory. 
Now $\Sigma$ can be integrated out, and this results in a constraint 
\begin{equation}
\sum_{i=1}^nq_iY_i-2t = 0. 
\end{equation}
Let us consider a simple case; all $q_i$ except $q_n=-q$ are positive and $\sum_{i=1}^nq_i<0$. 
Then the twisted superpotential, after integrating out $\Sigma$ is 
\begin{equation}
\tilde{W} = \mu\sum_{i=1}^{n-1}e^{-Y_i}
 +\mu e^{\frac2{q}t}\prod_{i=1}^{n-1}e^{-\frac{q_i}{q}Y_i}. 
\end{equation}

Let us define $u_i=e^{-\frac1qY_i}$. 
Then $\tilde{W}$ is 
\begin{equation}
\tilde{W} = \mu\sum_{i=1}^{n-1}u_i^q+\mu e^{\frac2qt}\prod_{i=1}^{n-1}u_i^{q_i}. 
   \label{tildeW}
\end{equation}
In the UV limit ($t_R\to-\infty$), the second term in the RHS vanishes. 
Naively, this limit of the GLSM is expected to be described by an $N=2$ SCFT which is specified by 
$\tilde{W}|_{t\to-\infty}$ (up to an orbifolding)\footnote{
The natural variables are $Y_i$, not $u_i$ itself. So the UV limit is not a product of minimal models. 
}. 
The second term of the RHS of (\ref{tildeW}) would indicate a perturbation by a 
tachyon background which grows 
at low energy $(t_R\to+\infty)$. 
In fact, the $U(1)$ charge shows that $\prod_{i=1}^{n-1}u_i^{q_i}$ is relevant, since 
$\sum_i^{n-1}q_i<q$ by assumption. 

Therefore, one may naturally expect that the GLSM would describe an RG flow which is induced by a 
tachyon perturbation. 
The endpoint of the RG flow is expected to describe a background in which the tachyon condenses, 
and, as a result, the target space is drastically deformed by the tachyon condensation, 
even at the level of topology. 
This approach to closed string tachyon condensations has been applied to non-supersymmetric 
orbifolds \cite{Vafa}. 
See also a review paper \cite{review}. 

Note that for another choice $\sum_iq_i>0$, $\prod_{i=1}^{n-1}u_i^{q_i}$ is irrelevant, 
while for this choice $t_R\to-\infty$ is the IR limit, 
so the irrelevant perturbation decays as $\Lambda$ decreases, 
as it should be the case. 
However, in this case it is not obvious which tachyons would induce this RG flow.

\vspace{5mm}

\subsection{Promoting $t$ to field}

\vspace{5mm}

The off-shell analysis based on an RG flow seems to be in conflict with the c-theorem, when one would 
like to discuss a generic tachyon which is not localized in a non-compact space. 
In such a case, for example a tachyon localized in a compact space \cite{Suyama2}, 
there is no reason for the 
c-theorem not to be applicable. 
Then the central charge must decrease along the RG flow induced by the tachyon, and therefore the 
consistency of string theory becomes suspicious. 
One may think that the endpoint of such a tachyon condensation would be a non-critical string theory, 
which is possible when a non-trivial dilaton background is induced by the tachyon condensation. 
One such mechanism was discussed in \cite{MIT}.

Some of recent researches \cite{Suyama2}\cite{MIT}\cite{Suyama1} 
focus on on-shell processes of closed string 
tachyon condensations. 
Since they are time-dependent processes, understanding of them is usually difficult. 
However, if one is able to analyze an on-shell process of tachyon condensation, the endpoint of 
the condensation can be identified as a state in string theory with which one had started. 

In this paper, we would like to show that it is possible to construct an exact CFT background in which 
a tachyon varies along a {\it spatial} direction, not the temporal one. 
It would be very interesting if the spatial direction can be Wick-rotated, enabling us to discuss the 
corresponding time-dependent processes. 

The construction is a rather straightforward generalization of \cite{Vafa}; 

it is achieved by promoting the parameter $t$ to a twisted chiral 
superfield $Y$, that is, 
the Lagrangian is 
\begin{equation}
L = \frac1{2\pi}\Bigl[ \int d^4\theta\ \Bigl( \sum_{i=1}^n\bar{\Phi}_ie^{2q_iV}\Phi_i
 -\frac1{2e^2}\bar{\Sigma}\Sigma -k\bar{Y}Y \Bigr) -\int d^2\tilde{\theta}\ Y\Sigma 
 -\int d^2\bar{\tilde{\theta}}\ \bar{Y}\bar{\Sigma} \Bigr]. 
    \label{promoted}
\end{equation}
Then the shift (\ref{t-shift}) due to the RG flow becomes an additive renormalization of $Y$, making 
the total Lagrangian kept intact. 
The dual theory of (\ref{promoted}) is almost the same as before, and in particular, the twisted 
superpotential is (\ref{tildeW}), with $t$ replaced with $Y$. 
The term $e^{\frac2qY}\prod_{i=1}^{n-1}u_i^{q_i}$ in the twisted superpotential can be regarded as a 
``dressed'' operator, which varies in $Y$-direction. 
Therefore, the target space of (\ref{promoted}) should describe an on-shell background in which 
a tachyon varies along the $Y$-direction. 
In one side of $Y$-direction, (\ref{promoted}) describes a background without tachyon, and in the 
other side it describes an endpoint of the tachyon condensation. 

Note that the GLSM (\ref{promoted}) is a mirror version of the one discussed in \cite{HoriKapustin}. 
See also \cite{twistedcircle} for its application to tachyon condensation. 

\vspace{5mm}

One advantage of introducing $Y$ is that the whole process of tachyon condensation can be described 
by one CFT which is obtained by taking the IR limit of the GLSM. 
In arguments relying on RG flows, one can only suggest that a CFT, which is an IR fixed point of a 
flow, would be a possible endpoint of a tachyon condensation. 
In our case, a tachyon varies in a spatial direction, keeping the on-shell condition, since our model 
is a CFT as a whole, not only in an asymptotic region. 
Therefore, the corresponding state with a non-trivial tachyon is connected to the original tachyonic 
vacuum via a physical process in string theory. 
It is also interesting that $Y$, along which the tachyon varies, looks like the Liouville field. 
This may have something to do with our previous works \cite{Suyama1}\cite{Suyama2}. 

In addition, $N=2$ superconformal symmetry is realized in our model. 
$N=2$ algebra has a $U(1)$ current which is a combination of $U(1)_V$ and $U(1)_A$ symmetries in the 
GLSM. 
By regarding the superfields in two dimensions as those obtained from four-dimensional ones via 
the dimensional reduction, $U(1)_V$ and $U(1)_A$ both have their origins in four dimensions; 
$U(1)_V$ comes from the R-symmetry, and $U(1)_A$ is the rotation in the plane which is dimensionally 
reduced. 
Since $U(1)_A$ symmetry has a chiral anomaly in the GLSM (\ref{Vafa}) when $\sum_iq_i\ne0$, 
it is not obvious 
whether $N=2$ symmetry is realized even at the IR limit, which seems to be generally expected. 

When the FI-parameter $t$ is promoted to $Y$, the chiral anomaly can be canceled. 
The chiral anomaly appears for the following transformation of fermions 
\begin{equation}
\psi_\pm \to e^{\mp ia}\psi_\pm, 
\end{equation}
under which the path-integral measure is not invariant, and the variation of the measure is 
\begin{equation}
{\cal D}\psi{\cal D}\bar{\psi} 
\to {\cal D}\psi{\cal D}\bar{\psi}\exp\Bigl[ -\frac{iq}{\pi}\int d^2x\ av_{01} \Bigr], 
\end{equation}
where $q$ is the gauge charge of $\psi_\pm$. 
All fermions in $\Phi_i$ gives this variation with appropriate charges. 
All of them can be canceled if the chiral transformation is accompanied by the shift of $Y$, 
\begin{equation}
Y \to Y+ia\sum_{i=1}^nq_i,  
\end{equation}
since the twisted superpotential in terms of component fields includes 
\begin{equation}
\frac1{\pi}\int d^2x\ y_Iv_{01},
\end{equation}
where $y=y_R+iy_I$ is the lowest component of $Y$. 
Note that the above term indicates that $Y$ is periodic with period $2\pi i$. 

In fact, one can construct generators of $N=2$ algebra acting on a subspace of the full Hilbert space 
of the CFT, according to \cite{SCA1}\cite{SCA2}\cite{HoriKapustin}. 
The details of construction of the generators are summarized in Appendix \ref{N=2}. 
The ambiguity of the choice of the $U(1)$ current mentioned there can be fixed for the GLSM 
(\ref{promoted}). 
Since the GLSM can be described by a non-linear sigma model in the IR limit, the $U(1)$ current of the 
$N=2$ algebra should be just a fermion number current (contribution from $Y$ can include a shift in 
$y_I$-direction). 
Therefore, one should choose $p_i=0$, since the lowest component of $J_2^i$ generates a transformation 
of $\phi_i$ which may act non-trivially on the target space. 
As a result, the central charge of this CFT is 
\begin{equation}
c=3(n-1)+3\Bigl( 1+k\gamma^2 \Bigr), 
\end{equation}
as expected from the dimensionality of the 
target space (plus linear dilaton background).

\vspace{1cm}

\section{Examples} \label{example}

\vspace{5mm}

In this section, we show explicit examples of GLSM's describing space-dependent tachyon backgrounds. 

The first example is (\ref{promoted}) with $n=1$. 
This is nothing but a mirror dual of the GLSM studied in \cite{HoriKapustin}, in which 
it has been shown that this GLSM is equivalent (mirror dual) to both the $N=2$ Liouville theory and 
$SL(2,{\bf R})/U(1)$ coset. 
The latter CFT describes a Euclidean black hole in two dimensions \cite{BH}, and its geometry is 
a semi-infinite cigar-like one. 
The shape of this target space can be seen from the classical vacuum manifold determined by 
\begin{equation}
|\phi|^2 = 2y_R. 
   \label{cigar}
\end{equation}
For simplicity, we have chosen $q=1$. 
For a positive $y_R$, (\ref{cigar}) has a solution, and its shape is locally a cylinder. 
For a negative $y_R$, however, there is no supersymmetric solution. 
Therefore the cylinder in the $y_R>0$ region is ``capped'' at $y_R=0$, resulting in the cigar 
geometry. 

The twisted superpotential of the dual theory is 
\begin{equation}
\tilde{W} = e^{-2Y}. 
\end{equation}
This is the $N=2$ Liouville potential. 
This can also be regarded as a tachyon vertex operator dressed by the Liouville field. 
If this operator is interpreted as a space-dependent tachyon background, it represents a tachyon 
growing toward $Y\to-\infty$. 
Then the original GLSM, which is a non-linear sigma model on the cigar, indicates that the target 
space ``disappears'' where tachyon condenses. 

The situation would be similar for a general $n$ case with $q_i>0$. 
The target space is asymptotically $WCP_{q_1,\cdots,q_n}\times {\bf R}\times S^1$ which is terminated 
at $y_R=0$. 
The mirror dual theory has a tachyon background which grows where the target space disappears. 

\vspace{5mm}

The second example is the GLSM with $n=2$ and $q_1q_2<0$. 

For simplicity, let $q_1=1$, $q_2=-2$. 
The corresponding twisted superpotential is 
\begin{equation}
\tilde{W} = e^{-Y_1}+e^{Y}e^{-\frac12Y_1}. 
\end{equation}
This can be regarded as describing a small tachyon perturbation if $Y\to-\infty$. 
Since there is a Liouville potential for $Y_1$, the region with large negative $Y_1$ is not accessible 
from the asymptotic $Y_1\to+\infty$ region. 
In the original variable, $Y_1\to-\infty$ corresponds to $\Phi_1\to0$ \cite{HoriVafa}. 

The vacuum manifold is determined by 
\begin{equation}
|\phi_1|^2-2|\phi_2|^2 = 2y_R. 
   \label{tube}
\end{equation}
We fix the gauge by imposing $\phi_1\ge0$. 
For the case $y_R<0$, this condition is not appropriate since $\phi_1=0$ is a possible solution of 
(\ref{tube}). 
However, $\phi_1=0$ is not physically relevant in this case, as mentioned above, so the gauge fixing 
condition $\phi_1\ge0$ would make sense. 

For $y_R<0$, the target space is a half of a one-sheeted hyperboloid, topologically a cylinder and 
the radius of the cylinder $|\phi_2|$ reduces as $\phi_1$ decreases. 
The minimum value of the radius becomes small as $|y_R|$ becomes small. 
The tachyon vertex $e^{-\frac12Y_1}$ in (\ref{tube}) is localized around $\phi_1=0$ and its amplitude 
$e^{Y}$ grows with $Y$. 
On the other hand, for $y_R>0$, the target space is instead a half of a two-sheeted hyperboloid, 
topologically a plane. 
Therefore, this can be understood that the thin part of the one-sheeted hyperboloid is pinched off by 
a tachyon condensation, which is the phenomenon discussed in \cite{Silverstein}. 

It seems very interesting 
that the GLSM discussed here is the same as the one employed in \cite{twistedcircle} 
for application to twisted circle geometry, but it is analyzed here with a different gauge choice, 
and $Y$ direction is regarded here as a physical coordinate.

\vspace{1cm}

\section{GLSM with superpotential} \label{W}

\vspace{5mm}

So far, we have discussed GLSM's without superpotential and their application to closed string 
tachyon condensations. 
In this section, we turn our attention into GLSM's with superpotential. 

\vspace{5mm}

One example which we would like to discuss is the GLSM (\ref{general}) with $n=2$, and the 
superpotential is 
\begin{equation}
W = \Phi_1^{n_1}\Phi_2^{n_2}, 
   \label{eg1}
\end{equation}
where $n_1,n_2>3$ are positive integers. 
This superpotential must be compatible with the gauge symmetry, which implies 
\begin{equation}
n_1q_1+n_2q_2 = 0. 
   \label{gaugeinv}
\end{equation}
For definiteness, we assume $q_1>0>q_2$ and $q_1+q_2<0$. 

The D-term condition is 
\begin{equation}
q_1|\phi_1|^2+q_2|\phi_2|^2 = 2y_R. 
\end{equation}
For $y_R<0$, $\phi_2$ cannot vanish, while for $y_R>0$, $\phi_1$ is non-zero. 
There is also the F-term condition which imposes $\phi_1=0$ for $y_R<0$ and $\phi_2=0$ for $y_R>0$. 
Summarizing, 
\begin{eqnarray}
\phi_1=0, &\hspace{5mm}& q_2|\phi_2|^2 = y_R<0, \\
\phi_2=0, &\hspace{5mm}& q_1|\phi_1|^2 = y_R>0. 
\end{eqnarray}
Then the vector superfield gets massive by the Higgs mechanism. For $y_R<0$, for example, 
\begin{equation}
\bar{\Phi}_2e^{2q_2V}\Phi_2 = |\Phi_1|^2+2q_2|\Phi_2|^2V+q_2|\Phi_2|^2V^2, 
\end{equation}
where the second term of the RHS cancels the term coming from the twisted superpotential. 
$\Phi_1$ and $Y$ remains massless. 
The superpotential for $\Phi_1$ is now 
\begin{equation}
W = a'\Phi_1^{n_1}. 
   \label{minimal}
\end{equation}
Therefore, around a vacuum with $y_R<0$, the GLSM describes, in the low energy limit, a CFT which 
consists of an $N=2$ minimal model specified by (\ref{minimal}) and a free field, 
with a possible linear dilaton background. 
Similarly, around a vacuum with $y_R>0$, the low energy limit is a product of another $N=2$ minimal 
model and a free field. 

Since there remains a residual gauge symmetry, the minimal model is in fact an orbifold. 
Consider the $y_R<0$ case. 
In this case, $\phi_2$ has a non-zero vev, so the gauge symmetry is broken to a discrete symmetry 
whose action on $\Phi_1$ is 
\begin{equation}
\Phi_1 \to e^{2\pi i\frac{q_1}{q_2}}\Phi_1 = e^{-2\pi i\frac{n_2}{n_1}}\Phi_1. 
\end{equation}
If $n_1,n_2$ are coprime to each other, the minimal model is orbifolded by ${\bf Z}_{n_1}$, otherwise 
the orbifold group is a subgroup of ${\bf Z}_{n_1}$. 

\vspace{5mm}

The dual theory provides a more detailed description of the model. 
The mirror dual of a GLSM with a superpotential was discussed in \cite{HoriVafa}. 
Assuming that their prescription is applicable to our case, the dual superpotential is 
\begin{equation}
\tilde{W} = X_1^{|q_2|}+e^{\frac2{|q_2|}Y}X_1^{q_1}, 
   \label{mirror}
\end{equation}
which is suitable for $Y\to-\infty$ limit, and 
\begin{equation}
\tilde{W} = X_2^{q_1}+e^{-\frac2{q_1}Y}X_2^{|q_2|}, 
\end{equation}
for $Y\to+\infty$ limit. 

The difference from the cases in section \ref{GLSM} 
is that the fundamental variables are $X_{1,2}$, not 
$\log X_{1,2}$. 
So the dual theory is really a minimal model, with a relevant (irrelevant) perturbation for 
$y_R\to-\infty$ ($y_R\to+\infty$), respectively (recall that 
$y_R\to-\infty$ is the IR limit when $q_1+q_2<0$). 
The minimal model appeared in (\ref{mirror}) is not different from (\ref{minimal}), since the 
gauge invariance condition (\ref{gaugeinv}) is solved as follows, 
\begin{equation}
q_1 = m n_2, \hspace{5mm} q_2 = -m n_1, 
\end{equation}
where $m$ can be absorbed by a rescaling of the gauge coupling $e$. 

Recall our assumption $q_1+q_2<0$. 
This implies $n_1>n_2$ which means that the central charge coming from the minimal model decreases 
by the tachyon condensation, as it should be the case. 

\vspace{5mm}

One can construct an $N=2$ superconformal algebra acting on a subspace of the Hilbert space. 
See Appendix \ref{N=2} for the details. 
The presence of such algebra strongly suggests that the IR limit of the GLSM is actually 
an $N=2$ SCFT with 
central charge 
\begin{equation}
c = 3(1-p_1)+3(1-p_2)+(-3)+3\Bigl( 1+k\gamma^2 \Bigr), 
\end{equation}
where $p_1,p_2$ are the $U(1)_V$ charges of $\Phi_1,\Phi_2$, respectively. 

It should be emphasized that the construction of the superconformal algebra does not depend on a 
particular limit, like $Y\to\pm\infty$. 
This implies that the IR limit is a CFT describing the following background: 
The background has one spatial direction corresponding to $y_R$. 
There is a non-trivial tachyon field which is exponentially growing in $y_R\to-\infty$ region, and 
looks like a massive field and is exponentially damped in $y_R\to+\infty$ region. 
The existence of such a CFT is very interesting since this can be regarded as an $\alpha'$-exact 
version of a solution of the equation of motion of the classical string theory, which has been 
discussed in \cite{Suyama1}\cite{YZ1}. 

Recall that the central charge coming from the minimal model part decreases after a tachyon 
condensation. 
Since total central charge must be the same for both $y_R\to\pm\infty$ region, which are parts of the 
same theory, the central charge coming from $Y$ field must increase after the tachyon 
condensation. 
This means that the dilaton gradient varies as tachyon condenses, which has been observed in the 
analysis of a solution of a low energy effective theory 
\cite{Suyama1}\cite{YZ1}\cite{MIT} and also in \cite{Suyama2}. 

\vspace{5mm}

Another example is the GLSM with 
\begin{equation}
W = \Phi_{0}G(\Phi_1,\cdots,\Phi_n), 
\end{equation}
where $G(x_1,\cdots,x_n)$ is a quasi-homogeneous polynomial satisfying 
\begin{equation}
G(\lambda^{q_1}x_1,\cdots,\lambda^{q_n}x_n) = \lambda^{-q_0}G(x_1,\cdots,x_n), 
\end{equation}
for $\lambda\in{\bf C}^*$. 
$G(x_1,\cdots,x_n)$ is also required to be regular, that is, $G=0$ and $\partial_iG=0$ have no 
common solution. 
The charges are assumed to satisfy 
\begin{equation}
q_1,\cdots,q_n<0<q_0, \hspace{5mm} \sum_{i=1}^n|q_i| > q_0. 
\end{equation}
The IR limit in $y_R\to-\infty$ region is an analogy of the Calabi-Yau phase \cite{GLSM}, that is, 
that region is described by a non-linear sigma model with target space $M\times{\bf R}$ where $M$ is 
a hypersurface $G=0$ in $WCP_{q_1,\cdots,q_n}$. 
On the other hand, $y_R\to+\infty$ region is described by a LG orbifold with superpotential 
$W=G(\Phi_1,\cdots,\Phi_n)$. 
Since one can construct an $N=2$ superconformal algebra in the previous example, it would be 
reasonable to expect that there is an $N=2$ SCFT which interpolates the above two theories. 
This CFT may describe a decay of $M$ into an LG orbifold via a tachyon condensation. 
This may be an example in which a decay of a compact manifold would result in a final state different 
from ``nothing''. 

The dual superpotential is, however, not what would be expected. 
One would obtain the following superpotential 
\begin{equation}
\tilde{W} = \sum_{i=1}^nX_i^{q_0}+e^{-\frac2{q_0}Y}\prod_{i=1}^nX_i^{|q_i|}. 
\end{equation}
This is suitable for $Y\to+\infty$ limit, but this limit should be described by the different 
superpotential, as mentioned above. 
It is very interesting to understand this situation better.

\vspace{1cm}

\section{$U(1)_V$ charges, central charge and gauge symmetry} \label{charge}

\vspace{5mm}

We have discussed various GLSM's in which $\gamma=\sum_iq_i\ne0$, 
and the FI-parameter promoted to a twisted 
chiral superfield. 
This kind of models has some properties which are absent in GLSM's with $\gamma=0$. 

It might seems strange that the central charge (\ref{center}) depends on the $U(1)_V$ charges $p_i$ of 
$\Phi_i$. 
The definition 
of $p_i$ is ambiguous, since $U(1)_V$ current can be modified by adding the gauge current. 
As a result, $p_i'=p_i+mq_i$ can also be regarded as $U(1)_V$ charges. 
However, this modification does not keep the central charge (\ref{center}) fixed, unless $\gamma=0$. 

One might think that this is related to the fact that the gauge current 
\begin{equation}
j_g = \sum_{i=1}^nq_i\Bigl[ 2i{\cal D}_-\bar{\phi}_i\phi_i-\bar{\psi}_{i,-}\psi_{i,-} \Bigr], 
\end{equation}
which acts on the $\bar{Q}_+$-closed subspace, is not a primary field, 
\begin{equation}
T(x)j_g(0) \sim \frac{i\gamma}{(x^-)^3}-\frac1{(x^-)^2}j_g(0)-\frac1{x^-}\partial_-j_g(0). 
\end{equation}
It is known that the presence of the 
$(x^-)^{-3}$ term is a signal of the appearance of the mixed anomaly, 
\begin{equation}
\partial_\mu j_g^\mu = a R^{(2)}, 
\end{equation}
where $R^{(2)}$ is the scalar curvature of the worldsheet. 
However, contributions to 
the coefficient $a$ comes not only from $\gamma$ but also from a coefficient of $(x^+)^{-3}$ 
term of the $\bar{T}\bar{j}_g$ OPE. 
By exchanging $-$ and $+$, one can similarly construct the $N=2$ algebra acting on $\bar{Q}_-$-closed 
subspace of the Hilbert space, and the OPE's of them are the same as those we have discussed. 
In particular, the coefficient of $(x^+)^{-3}$ in the $\bar{T}\bar{j}_g$ OPE is $i\gamma$. 
Since $a$ is proportional to the difference of these coefficients, it has been shown that the mixed 
anomaly is absent in our GLSM. 
Note that, of course, the zero mode of $j_g$ commutes with the generators of the $N=2$ algebra, so 
the gauge symmetry is preserved, although it is not promoted to an affine symmetry. 

If $\gamma=0$, $j_g$ is primary. 
Moreover, the following modified operators 
\begin{eqnarray}
T_\beta &=& T+\frac i4\beta\partial j_g, \\
G_\beta &=& G, \\
\bar{G}_\beta &=& \bar{G}+i\beta\sum_{i=1}^nq_i\partial(\phi_i\bar{\psi}_i), \\
j_\beta &=& j-\frac12\beta j_g,
\end{eqnarray}
form the $N=2$ algebra with the same central charge. 
Therefore, the ambiguity of the choice of $U(1)_V$ charges is absent in the $\gamma=0$ case, in the 
sense that all choices provide the same central charge. 

In fact, the dependence of such an ambiguity of central charge already 
exists in ungauged chiral model. 
In \cite{SCA1}, it is shown that a two-dimensional theory with the Lagrangian 
\begin{equation}
L = \int d^4\theta\ \sum_{i=1}^n\bar{\Phi}_i\Phi_i+\int d^2\theta\ W(\Phi)
    \label{chiral}
\end{equation}
reduces in the IR limit to an $N=2$ SCFT with central charge 
\begin{equation}
c = 3\sum_{i=1}^n(1-2\alpha_i), 
\end{equation}
where $\alpha_i$ are determined by 
\begin{equation}
W = \sum_{i=1}^n\alpha_i\Phi_i\partial_iW(\Phi). 
   \label{euler}
\end{equation}
Therefore, if (\ref{chiral}) has a $U(1)$ symmetry, that is, $W$ also satisfies 
\begin{equation}
0 = \sum_{i=1}^nq_i\Phi_i\partial_iW(\Phi),
\end{equation}
then $\alpha_i$ can be determined only up to $q_i$, and the central charge cannot be fixed. 

To see the origin of this ambiguity, consider the following superpotential, 
\begin{equation}
W_{\xi,m} = \Phi_1^{n_1}\Phi_2^{n_2}+\xi\Phi_1^{m_1}\Phi_2^{m_2}. 
\end{equation}
For this superpotential with $\xi\ne0$, (\ref{euler}) uniquely determines $\alpha_i$, unless 
$n_1m_2=n_2m_1$. 
Then the central charge depends on $m_i$, but not on $\xi$. 
One can take $\xi\to0$ limit while keeping the central charge fixed, so various CFT's with different 
central charges degenerate at $W=\Phi_1^{n_1}\Phi_2^{n_2}$ which we have discussed in the previous 
section. 
Therefore, the IR limit would not be uniquely determined for such a ``degenerate'' superpotential, 
and there would be {\it the} $U(1)_V$ charges $\alpha_i$ (or $p_i$ in our notation) which should be 
distinguished from a $U(1)_V$ charges dictating the transformation property of fields. 

As mentioned above, any GLSM with a superpotential may have such an ambiguity, since the 
superpotential must preserve the gauge symmetry. 
It is very interesting to know how to fix this ambiguity.

\vspace{1cm}

\section{Discussion} \label{discuss}

\vspace{5mm}

We have discussed various GLSM's in which the FI-parameter is promoted to a twisted chiral 
superfield $Y$, 
and its application to closed string tachyon condensations. 
The presence of $Y$ enables us to cancel a possible chiral anomaly, which results in the possibility 
to have an $N=2$ SCFT in the IR limit. 
It is remarkable that any RG flow described by a GLSM with an FI-parameter can be used to construct an 
$N=2$ SCFT. 
Since it is conformal, it can be used for a background of string theory, which is $\alpha'$-exact 
solution of the equations of motion at the tree level. 
From the mirror description, at least some of them can be regarded as describing 
a background in which a tachyon 
grows along the $y_R$-direction. 
Therefore, the GLSM accomplishes the construction of an $\alpha'$-exact description of an on-shell 
tachyon condensation from the corresponding worldsheet RG flow. 
By using 
such GLSM's, we have described tachyon condensations in which a part of a target space disappears 
or the topology of the target space changes. 
In another cases, we have shown that the dilaton gradient varies as a tachyon condenses, when the 
condensation decreases the central charge of a part of the system. 
The latter phenomenon has been observed in 
\cite{Suyama1}\cite{YZ1}\cite{MIT} at the level of the spacetime effective theory, 
and in this paper we have shown that this is indeed the case also in the level of $\alpha'$-exact 
solutions. 
It is also interesting that the RG scale (FI-parameter in GLSM) is related to the 
field $Y$ in order to 
obtain the consistent background for tachyon condensations, and the $Y$ field realy looks like the 
Liouville field. 
This may be understood as an explicit realization of the description of tachyon condensations 

proposed in \cite{Suyama2}. 

It may seem curious that the explicit expression of the energy-momentum tensor (\ref{EM}) 
shows a simple 
linear dilaton background for $y$, although we claimed that the dilaton gradient varies along $y_R$. 
In fact, our claim is that a non-trivial dilaton appears after some degrees of freedom, becoming 
massive, are integrated out. 
Note that the contribution to the 
central charge from the vector superfield $V$ is $-3$, which is the right 
amount to cancel the contribution from a chiral superfield with zero $U(1)_V$ charge. 
So the Higgs mechanism and the successive integrating out massive fields does not change the total 
central charge, if the $U(1)_V$ charge of fields are suitably chosen. 
For example, If one chooses $p_1=\frac2{n_1}, p_2=0$ for the case (\ref{eg1}), then in the 
$y_R\to-\infty$ 
region where $\Phi_2$ becomes massive, the contributions to the central charge from $\Phi_2$ and $V$ 
cancel each other. 
So we can simply ignore those fields, and therefore the dilaton gradient is the one appears in the 
energy-momentum tensor (\ref{EM}). 
Then what happens in the $y_R\to+\infty$ region? 
As mentioned in section \ref{charge}, 
one cannot redefine the $U(1)_V$ charges as $p_1=0,p_2=\frac2{n_2}$ by 
using the gauge symmetry. 
Instead, one can regard the $N=2$ algebra for $(p_1,p_2)=(\frac2{n_1},0)$ as the algebra for 
$(p_1,p_2)=(0,\frac2{n_2})$ ``twisted'' by the gauge current. 
For example, 
\begin{equation}
T_{p_1=\frac2{n_1},p_2=0} = T_{p_1=0,p_2=\frac2{n_2}}+\frac i4\beta\partial_-j_g, 
\end{equation}
where $\beta=\frac2{n_1q_1}=-\frac2{n_2q_2}$. 
The central charge coming 
from $\Phi_1$ through $T_{p_1=0,p_2=\frac2{n_2}}$ is canceled by $V$, but $j_g$ 
also contributes to the central charge, due to the presence of the $(x^-)^{-3}$ term, and this 
part would be absorbed to the Liouville part 
by a non-trivial field redefinition, resulting in a dilaton gradient 
different from that in (\ref{EM}). 

It is worth emphasizing again that for any RG flow described by a GLSM one can construct an $N=2$ 
SCFT in which the RG scale is replaced with a spatial coordinate. 
Therefore, in principle, we have obtained various CFT's 
describing tachyon condensations of various kinds. 
Since it is on-shell backgrounds, it is indeed possible to be realized in string theory. 
Several examples were discussed in this paper. 
It is expected that a systematic research of these CFT's and the corresponding tachyon condensations 
would provide a deeper understanding of closed string tachyon condensations. 
In particular, it would help understanding which kind of tachyons makes the target spaces 
disappeared and which are not. 
Note that, as mentioned at the end of in section \ref{W}, 
not all CFT's obtained from GLSM describe target space 
dynamics induced by tachyon condensations. 
It is very important to clarify this issue. 

The relation between a two-dimensional black hole \cite{BH} and a tachyon condensation, mentioned in 
section \ref{example}, seems to be 
interesting. 
Naively, by Wick-rotating $y_I$-direction, not $y_R$, one might obtain the Lorentzian black hole 
background, and the region where the tachyon condenses is behind the horizon. 
Therefore, it is tempting to guess that an inhomogeneous tachyon condensation would result in a 
formation of a black hole. 
Then a homogeneous tachyon condensation might be understood that the whole target space falls 
inside the 
horizon. 
Another interesting point of this relation is that the two-dimensional black hole has a matrix model 
description \cite{KKK}. 
It is very interesting if one can use such a matrix model technique to analyze a closed string tachyon 
condensation non-perturbatively. 
A non-perturbative analysis of closed string tachyon condensation is very important since a typical 
endpoint of the condensation would be strongly coupled. 
A relation between tachyon condensation and two-dimensional black hole was already mentioned in 
\cite{Olsson}. 

So far, we have discussed static space-dependent tachyon background. 
To study a time evolution of a tachyon condensation, one should perform the Wick rotation of the 
spacelike Liouville direction into the timelike one. 
There may be another way to extract a time-dependent process from our CFT. 
Let us discuss the spacetime effective action employed in \cite{Suyama1}\cite{YZ1}.  
To obtain a space-dependent solution, one would make an ansatz 
\begin{equation}
ds^2 = e^{A(y)}\eta_{\mu\nu}dx^\mu dx^\nu+dy^2, \hspace{5mm} \Phi=\Phi(y), \hspace{5mm} 
T=T(y). 
\end{equation}
Then the equations of motion reduce to the following ones, 
\begin{eqnarray}
\frac12(T')^2-V(T) &=& 2(\Phi')^2, \\
\Phi'' &=& \frac12(T')^2, \\
A &=& \mbox{const.} 
\end{eqnarray}
At a local minimum below zero, the tachyon $T$ can stop its rolling, and the dilaton is linear in 
$y$-direction. 
Consider a geodesic motion in this background in the Einstein frame. 
The geodesic equation is 
\begin{equation}
\frac{d^2y}{dt^2} = -\frac12\phi'\Bigl[ 1-\Bigl( \frac{dy}{dt} \Bigr)^2 \bigr], 
\end{equation}
where $\phi=-\frac4{D-1}\Phi$ and $t=x^0$. 
This indicates that all test particles are accelerated toward the positive $y$-direction, 
and its speed approaches the speed of light. 
The monotonicity of the geodesics seems to be related to the monotonicity 
of the RG flow behind the tachyon 
condensation. 
It may be interesting if the target space which an observer moving along the geodesic sees is a 
Penrose limit of the above solution. 
It is known that a Penrose limit of any solutions in string theory and M-theory preserves at least 
16 supersymmetries, so the properties of the endpoint of the condensation might be understood, 
although the corresponding background is strongly coupled and the metric becomes singular.

\vspace{2cm}

\begin{flushleft}
{\Large \bf Acknowledgments}
\end{flushleft}

\vspace{5mm}

I would like to thank S.Iso, T.Takayanagi and T.Tokunaga for valuable discussions. 
This work was supported in part by JSPS Research Fellowships for Young Scientists.

\newpage

\appendix

\hspace*{-5mm}{\bf \Large Appendix}

\section{Conventions} \label{convention}

\vspace{5mm}

Our conventions for superfields are obtained by the dimensional reduction of four-dimensional ones 
to two-dimensions. 
More specifically, we follow the conventions in cite{WessBagger}, reduce $1,2$-directions and rename 
$x^3$ as $x^1$. 
The super-derivatives are 
\begin{equation}
D_\pm = \frac{\partial}{\partial\theta^\pm}-2i\bar{\theta}^\pm\partial_\pm, \hspace{5mm} 
\bar{D}_\pm = -\frac{\partial}{\partial\bar{\theta}^\pm}+2i\theta^\pm\partial_\pm, 
\end{equation}
where $\partial_\pm=\frac12(\partial_0\pm\partial_1)$. 

The chiral superfield $\Phi$ in the component fields is 
\begin{eqnarray}
\Phi &=& \phi-i\theta^-\bar{\theta}^-\partial_-\phi-i\theta^+\bar{\theta}^+\partial_+\phi
 +\theta^+\theta^-\bar{\theta}^+\bar{\theta}^-\partial_+\partial_- \phi+2\theta^+\theta^-F \nonumber \\
& &+\sqrt{2}(\theta^-\psi_-+\theta^+\psi_+)
 -i\sqrt{2}\theta^+\theta^-(\bar{\theta}^-\partial_-\psi_+-\bar{\theta}^+\partial_+\psi_-). 
\end{eqnarray}
The vector superfield $V$ in the Wess-Zumino gauge is 
\begin{eqnarray}
V &=& \theta^+\bar{\theta}^+v_++\theta^-\bar{\theta}^-v_--\theta^-\bar{\theta}^+\sigma
 -\theta^+\bar{\theta}^-\bar{\sigma}-2\theta^+\theta^-\bar{\theta}^+\bar{\theta}^-D \nonumber \\
& & -2i\theta^+\theta^-(\bar{\theta}^-\bar{\lambda}_-+\bar{\theta}^+\bar{\lambda}_+)
 +2i\bar{\theta}^+\bar{\theta}^-(\theta^-\lambda_-+\theta^+\lambda_+). 
\end{eqnarray}
From this, one can obtain the following expression of $\Sigma$, 
\begin{eqnarray}
\Sigma 
&=& \bar{D}_+D_-V \nonumber \\
&=& \sigma+2i\theta^+\bar{\lambda}_+-2i\bar{\theta}^-\lambda_-+2\theta^+\bar{\theta}^-(D-iv_{01}) 
 \nonumber \\
& &-i\theta^+\bar{\theta}^+\partial_+\sigma+i\theta^-\bar{\theta}^-\partial_-\sigma
 -\theta^+\theta^-\bar{\theta}^+\bar{\theta}^-\partial_+\partial_-\sigma \nonumber \\
& &-2\theta^+\theta^-\bar{\theta}^-\partial_-\bar{\lambda}_+
 -2\bar{\theta}^+\bar{\theta}^-\theta^+\partial_+\lambda_-. 
\end{eqnarray}

To define twisted superfields, it is convenient to introduce the following involution $I$, 
\begin{equation}
I(\theta^-) = -\bar{\theta}^-, \hspace{5mm} I(\bar{\theta}^-) = -\theta^-, 
\end{equation}
which preserves the ordering of the Grassmann coordinates. 
$I$ exchanges $D_-$ and $\bar{D}_-$, while keeps $D_+$ and $\bar{D}_+$ fixed, so $I$ exchanges chiral 
superfields and twisted chiral superfields. 
The component expansion of the twisted superfield $Y$ is therefore, 
\begin{eqnarray}
Y &=& y+i\theta^-\bar{\theta}^-\partial_-y-i\theta^+\bar{\theta}^+\partial_+y
 -\theta^+\theta^-\bar{\theta}^+\bar{\theta}^-\partial_+\partial_- y-2\theta^+\bar{\theta}^-F_Y
 \nonumber \\
& &+\sqrt{2}(-\bar{\theta}^-\chi_-+\theta^+\bar{\chi}_+)
 +i\sqrt{2}\theta^+\bar{\theta}^-(-\theta^-\partial_-\bar{\chi}_+-\bar{\theta}^+\partial_+\chi_-). 
\end{eqnarray}
The kinetic term of $Y$ is obtained as the usual kinetic term of the chiral superfield $I(Y)$. 
That is, 
\begin{equation}
\int d^4\theta\ I(\bar{Y})I(Y) = -\int d^4\theta\ \bar{Y}Y. 
\end{equation}
The Grassmann integration measure is defined as $d^2\tilde{\theta}=d\bar{\theta}^-d\theta^+$, so 
the twisted superpotential can be rewritten as a superpotential, for example, 
\begin{equation}
\int d^2\tilde{\theta}\ Y\Sigma = -\int d^2\theta\ I(Y)I(\Sigma). 
\end{equation}

\vspace{1cm}

\section{N=2 algebra from GLSM} \label{N=2}

\vspace{5mm}

The Lagrangian of the GLSM discussed in this appendix is the following, 
\begin{equation}
L = \frac1{2\pi}\int d^4\theta\ \Bigl[ \sum_{i=1}^n\bar{\Phi}_ie^{2q_iV}\Phi_i-\frac1{2e^2}\bar{\Sigma}
 \Sigma-k\bar{Y}Y \Bigr]+\frac1{2\pi}\Bigl[ \int d^2\theta\ aW(\Phi)-\int d^2\tilde{\theta}\ Y\Sigma 
 +\mbox{h.c.}\ \Bigr], 
   \label{general}
\end{equation}
where $a$ is a constant. 
The equations of motion of the superfields are 
\begin{eqnarray}
\bar{D}_+\bar{D}_-(\bar{\Phi}_ie^{2q_iV}) &=& -2a\ \partial_iW(\Phi), \\
\frac1{2e^2}\bigl( D_+\bar{D}_-\Sigma+\bar{D}_+D_-\bar{\Sigma} \bigr) 
&=& \sum_{i=1}^n2q_i\bar{\Phi}_ie^{2q_iV}\Phi_i-2(\bar{Y}+Y), \\
k\bar{D}_+D_-\bar{Y} &=& -2\Sigma. 
\end{eqnarray}

The goal is to construct generators of $N=2$ superconformal algebra acting on a subspace of the full 
Hilbert space which preserves $\bar{Q}_+$. 
Such generators must be $\bar{Q}_+$-closed, that is, an operator $O$ such that $\{\bar{Q}_+,O\}=0$. 
Since $\bar{Q}_+=\bar{D}_+-2i\theta^+(\partial_0+\partial_1)$ on the superspace, 
the lowest component of a 
$\bar{D}_+$-closed superfield is $\bar{Q}_+$-closed. 
Therefore, a way to obtain the $N=2$ generators is to construct a $\bar{D}_+$-closed superfield whose 
lowest component is the $U(1)$ current of the $N=2$ algebra. 
The other generators can be obtained as higher components of the same superfield. 

\vspace{5mm}

\subsection{Classical analysis}

\vspace{5mm}

There are various $U(1)$ symmetries in (\ref{general}). 
Let us first consider a $U(1)$ symmetry acting on $\Phi$, suppressing the subscript. 
It is generated by $\partial_-\bar{\phi}\phi$, $\bar{\phi}_-\partial\phi$, $\bar{\psi}_-\psi_-$, where 
$\partial_-=\frac12(\partial_0-\partial_1)$, and their right-moving partners. 
Their gauge-invariant and supersymmetric extensions are as follows, 
\begin{eqnarray}
D_-\{e^{-2qV}\bar{D}_-(e^{2qV}\bar{\Phi})\}e^{2qV}\Phi 
&=& 4i(\partial_-\bar{\phi}-iv_-\bar{\phi})\phi+\cdots, 
    \label{current} \\
\bar{\Phi}e^{2qV}\bar{D}_-\{e^{-2qV}D_-(e^{2qV}\Phi)\} 
&=& 4i\bar{\phi}(\partial_-\phi+iv_-\phi)+\cdots, \\
\bar{D}_-(\bar{\Phi}e^{2qV})e^{-2qV}D_-(e^{2qV}\Phi) &=& 2\bar{\psi}_-\psi_-+\cdots. 
\end{eqnarray}
One can check that the second superfield cannot contribute to a $\bar{D}_+$-closed superfield, even by 
using the equations of motion, and what is discussed below is a linear combination of the the first 
and the third one. 
Let us define 
\begin{eqnarray}
J_1^i &=& \bar{D}_-(\bar{\Phi}_ie^{2q_iV})e^{-2q_iV}D_-(e^{2q_iV}\Phi_i), \\
J_2^i &=& D_-\bar{D}_-\Bigl( \bar{\Phi}_ie^{2q_iV}\Phi_i \Bigr). 
\end{eqnarray}
Note that $J_1^i+J_2^i$ is equal to (\ref{current}). 
Define 
\begin{equation}
J_\Phi = \sum_{i=1}^n(\alpha_iJ_1^i+\beta_iJ_2^i). 
\end{equation}
One can show that 
\begin{eqnarray}
\bar{D}_+J_\Phi 
&=& -\Sigma\bar{D}_-\Bigl[ \sum_{i=1}^n2\alpha_iq_i\bar{\Phi}_ie^{2q_iV}\Phi_i \Bigr] 
 -4a\sum_{i=1}^n\Bigl[ \alpha_iq_i\Phi_i\partial_iW(\Phi) \Bigr]D_-V \nonumber \\
& & +a\sum_{i=1}^n\Bigl[ +2\beta_iD_-(\Phi_i\partial_iW(\Phi))-2\alpha_i\partial_iW(\Phi)D_-\Phi_i
    \Bigr].  
\end{eqnarray}
Suppose that $a\ne0$. 
The second term of the RHS vanish if all $\alpha_i$ are equal to, say, $\alpha$. 
For a generic superpotential, this is the unique 
choice since there would exist only the gauge symmetry.  
Then, the third term vanishes if 
\begin{equation}
\sum_{i=1}^n \beta_i\Phi_i\partial_iW(\Phi) = \alpha W(\Phi). 
\end{equation}
If $\alpha=0$, then the solution is generically $\beta_i=q_i$. 
Otherwise, let $\beta_i=\frac \alpha2p_i$. 
Then 
\begin{equation}
\sum_{i=1}^n p_i\Phi_i\partial_iW(\Phi) = 2W(\Phi). 
   \label{U(1)_V}
\end{equation}
Therefore, $p_i$ are the $U(1)_V$ charges of $\Phi_i$. 

Now one obtains 
\begin{equation}
\bar{D}_+J_\Phi 
= -\alpha\Sigma\bar{D}_-\Bigl[ \sum_{i=1}^n2q_i\bar{\Phi}_ie^{2q_iV}\Phi_i \Bigr] 
\end{equation}
provided $a=0$ or 
\begin{equation}
\alpha_i = \alpha \hspace{5mm} (i=1,\cdots,n), \hspace{5mm} 
\beta_i = \left\{
\begin{array}{cc}
q_i, & (\alpha=0), \\ [2mm] \displaystyle{\frac\alpha2p_i}, & (\alpha\ne0), 
\end{array}
\right.
\end{equation}
where $p_i$ satisfy (\ref{U(1)_V}). 
Note that $J_\Phi$ with $\alpha=0$ is just the current for the global gauge symmetry, which is not 
appropriate for our purpose. 
In what follows, we assume non-zero $\alpha$, and normalize $J_\Phi$ so that $\alpha=1$. 
Note also that $p_i$ are not determined uniquely by (\ref{U(1)_V}) due to the global gauge symmetry. 

By using the equations of motion, one can show that 
\begin{equation}
J_c = \sum_{i=1}^n\Bigl[ J_1^i+\frac{p_i}2J_2^i \Bigr]
 -\frac1{2e^2}\Sigma\bar{D}_-D_-\bar{\Sigma}-kD_-\bar{Y}\bar{D}_-Y
\end{equation}
is $\bar{D}_+$-closed, classically. 

\vspace{5mm}

\subsection{Chiral anomaly}

\vspace{5mm}

At the quantum level, the current $J_c$ may be anomalous. 
The chiral anomaly actually appears from 
\begin{equation}
{\cal D}_-\bar{\phi}_i\phi_i, \hspace{5mm} \bar{\psi}_{i,-}\psi_{i,-}, 
\end{equation}
whose definitions are 
\begin{eqnarray}
{\cal D}_-\bar{\phi}_i\phi_i(0) 
&:=& \lim_{x\to0}\Bigl[ {\cal D}_-\bar{\phi}_i(x)e^{-iq_i\int_0^xd\xi^\mu v_\mu(\xi)}\phi_i(0) 
 +\frac1{2x^-} \Bigr] \nonumber \\
&=& :{\cal D}_-\bar{\phi}_i(0)\phi_i(0):+\frac i2q_iv_-(0)
 +\frac i2q_i\lim_{x\to0}\frac{x^+}{x^-}v_+(0).   \\
\bar{\psi}_{i,-}\psi_{i,-}(0) 
&:=& \lim_{x\to0}\Bigl[ \bar{\psi}_{i,-}(x)e^{-iq_i\int_0^xd\xi^\mu v_\mu(\xi)}\psi_{i,-}(0)
 +\frac i{x^-} \Bigr] \nonumber \\
&=& :\bar{\psi}_{i,-}(0)\psi_{i,-}(0):-q_iv_-(0)-q_i\lim_{x\to0}\frac{x^+}{x^-}v_+(0). 
\end{eqnarray}
Note that for the singular part of the OPE for $\phi_i$, the gauge-covariant one 
\begin{equation}
\bar{\phi}_i(x)\phi_i(0) \sim -\frac12e^{iq_i\int_0^xd\xi^\mu v_\mu(\xi)}\log(x^-x^+)
\end{equation}

should be used. 
Although the RHS goes to $-\frac12\log(x^-x^+)$ in the $x\to0$ limit, the terms of the form 
$x^n\log x$ cannot be neglected since their derivative may be singular at the limit. 
This should be the same for fermions, but in these cases the exponential term is always irrelevant. 

According to cite{HoriKapustin}, one obtains 
\begin{eqnarray}
\bar{D}_+J_1^i &=& -q_i\bar{D}_-\Sigma, \\
\bar{D}_+J_2^i &=& 0. 
\end{eqnarray}
Therefore,  
\begin{equation}
\bar{D}_+J_c = -\gamma\bar{D}_-\Sigma, \hspace{5mm} \gamma = \sum_{i=1}^nq_i.  
\end{equation}
Note that the current $J_c$ is $\bar{D}_+$-closed even at the quantum level if 
$\gamma=0$, which is relevant for a GLSM for a supersymmetric background of string 
theory. 

As the chiral anomaly can be canceled due to the presence of $Y$, $J_c$ can be modified by using $Y$ 
to recover the quantum $\bar{D}_+$-closedness. 
Consider a superfield $\bar{D}_-D_-\bar{Y}-D_-\bar{D}_-Y$ whose lowest component generates a shift of 
$y_I$. 

This satisfies 
\begin{equation}
\bar{D}_+\Bigl[ \bar{D}_-D_-\bar{Y}-D_-\bar{D}_-Y \Bigr] = \frac2k\bar{D}_-\Sigma. 
\end{equation}
This equation is valid at the quantum level. 
Therefore, the following current 
\begin{equation}
J = J_c+\frac{k\gamma}2(\bar{D}_-D_-\bar{Y}-D_-\bar{D}_-Y)
\end{equation}
is exactly $\bar{D}_+$-closed. 

\vspace{5mm}

\subsection{$N=2$ algebra}

\vspace{5mm}

The lowest component of the current $J$ is a candidate for the $U(1)$ current $j$ of $N=2$ algebra. 
It is obvious that $D_-J$, $\bar{D}_-J$ and $\left[D_-,\bar{D}_-\right]J$ are also $\bar{D}_+$-closed, 
and in fact, lowest components of them 
are the supercharges $G,\bar{G}$ and the energy-momentum tensor $T$, respectively. 
To check this statement, one has to calculate the OPE of them. 
It is a difficult tack in general, but in our case the situation is quite simple. 
Since the interaction terms only contribute to non-singular terms of OPE, one can calculate OPE 
of any operators by using (gauge-covariant) free propagators. 

One can show that 
\begin{equation}
j = -\frac12J\big|_1, \hspace{5mm} G = \frac1{2\sqrt{2}}D_-J\big|_1, \hspace{5mm} 
 \bar{G} = \frac1{2\sqrt{2}}\bar{D}_-J\big|_1, 
 \hspace{5mm} T = \frac1{16}\left[D_-,\bar{D}_-\right]J\big|_1, 
\end{equation}
form the $N=2$ algebra\footnote{
The algebra reduces to the well-known one, 
after the Wick rotation and a suitable rescaling of operators 
related to the conformal transformation from the 
cylinder to the plane.} cite{HoriKapustin} 
\begin{eqnarray}
T(x)T(0) &\sim& \frac {\frac c2}{(x^-)^4}-\frac2{(x^-)^2}T(0)-\frac1{x^-}\partial_-T(0), \\
T(x)G(0) &\sim& \frac{-\frac32}{(x^-)^2}G(0)-\frac1{x^-}\partial_-G(0), \\
T(x)\bar{G}(0) &\sim& \frac{-\frac32}{(x^-)^2}\bar{G}(0)-\frac1{x^-}\partial_-\bar{G}(0), \\
G(x)\bar{G}(0) &\sim& \frac{\frac23ic}{(x^-)^3}-\frac2{(x^-)^2}j(0)
 -\frac i{x^-}(2T(0)-i\partial_-j(0)), \\
T(x)j(0) &\sim& -\frac1{(x^-)^2}j(0)-\frac1{x^-}\partial_-j(0), \\
j(x)G(0) &\sim& -\frac i{x^-}G(0), \\
j(x)\bar{G}(0) &\sim& \frac i{x^-}\bar{G}(0), \\
j(x)j(0) &\sim& -\frac{\frac c3}{(x^-)^2}, 
\end{eqnarray}
with central charge 
\begin{equation}
c = \sum_{i=1}^n3(1-p_i)+(-3)+3\Bigl( 1+k\gamma^2 \Bigr), 
   \label{center}
\end{equation}
coming from $\Phi_i$, $V$ and $Y$, respectively. 

The explicit form of the generators are as follows, 
\begin{eqnarray}
j &=& \sum_{i=1}^n\Bigl[ -\bar{\psi}_i\psi_i-\frac{p_i}2(2i{\cal D}\bar{\phi}_i\phi_i
 -\bar{\psi}_i\psi_i) \Bigr] 
 +\frac i{e^2}\sigma\partial\bar{\sigma}+k\bar{\chi}\chi
 -ik\gamma\partial(\bar{y}-y), \\
G &=& \sum_{i=1}^n2i{\cal D}\bar{\phi}_i\psi_i
 -\frac{\sqrt{2}}{e^2}\sigma\partial\bar{\lambda}+2ik\bar{\chi}\partial y
 +ik\gamma\partial\bar{\chi}, \\
\bar{G} &=& \sum_{i=1}^n\bigl[ i(p_i-2){\cal D}\phi_i\bar{\psi}_i
 +ip_i\phi_i{\cal D}\bar{\psi}_i \bigr]
 +\frac{\sqrt{2}}{e^2}\lambda\partial\bar{\sigma}
 -2ik\chi\partial \bar{y}-ik\gamma\partial\chi, \\
T &=& \sum_{i=1}^n\Bigl[ 2{\cal D}\bar{\phi}_i{\cal D}\phi_i+\frac i2\bigl( \bar{\psi}_i{\cal D}\psi_i
 -{\cal D}\bar{\psi}_i\psi_i \bigr)+\frac i4p_i\partial\bigl( 2i{\cal D}\bar{\phi}_i\phi_i
 -\bar{\psi}_i\psi_i \bigr) \Bigr] \nonumber \\
& &  +\frac1{2e^2}(\partial\bar{\sigma}\partial\sigma-\sigma\partial^2\bar{\sigma})
 +\frac i{e^2}\lambda\partial\bar{\lambda}
 +2k\partial\bar{y}\partial y+\frac12k\gamma\partial^2(\bar{y}+y)
 +\frac i2k(\bar{\chi}\partial\chi-\partial\bar{\chi}\chi), 
   \label{EM}
\end{eqnarray}
where the subscript $-$ is suppressed.

\end{document}